\documentclass[preprint,showkeys]{revtex4-2}

\usepackage{graphicx}
\usepackage{dcolumn}
\usepackage{bm}
\usepackage{amsfonts}
\usepackage{amsmath}
\usepackage{braket}
\usepackage[utf8]{inputenc}
\usepackage{hyperref}
\usepackage{color,soul}

\providecommand{\keywords}[1]
{
  \small	
  \textbf{\textit{Keywords---}} #1
}

\begin{document}

\title{Path-Optimized Fast Quasi-Adiabatic Driving in Coupled Elastic Waveguides}

\author{Dong Liu$^{1,2}$}
\author{Yiran Hao$^{1,2}$}
\author{Jensen Li$^{1,3}$}
\email{jensenli@ust.hk}

\affiliation{$^{1}$Department of Physics, The Hong Kong University of Science and Technology, Hong Kong, P. R. China.}
\affiliation{$^{2}$School of Physical Science and Technology, Hunan University of Technology and Business, Changsha, China.}
\affiliation{$^{3}$Department of Engineering, University of Exeter, Exeter, United Kingdom.}

\begin{abstract}
Fast quasi-adiabatic driving (FAQUAD) is a central technique in shortcuts to adiabaticity (STA), 
enabling accelerated adiabatic evolution by optimizing the rate of change of a single control parameter. 
However, many realistic systems are governed by multiple coupled parameters, 
where the adiabatic condition depends not only on the local rate of change but also on the path through parameter space. 
Here, we introduce an enhanced FAQUAD framework that incorporates path optimization in addition to conventional velocity optimization, 
extending STA control to two-dimensional parameter spaces. 
We implement this concept in a coupled elastic-waveguide system, 
where the synthetic parameters—detuning and coupling—are controlled by the thicknesses of the waveguides and connecting bridges. 
Using scanning laser Doppler vibrometry, we directly map the flexural-wave field and observe adiabatic energy transfer along the optimized path in parameter space. 
This elastic-wave platform provides a versatile classical analogue for exploring multidimensional adiabatic control, 
demonstrating efficient and compact implementation of shortcut-to-adiabaticity protocols.
\end{abstract}

\keywords{Fast quasi-adiabatic driving, shortcut to adiabaticity, elastic metamaterials}

\maketitle

\section{Introduction}

The adiabatic theorem states that a time-dependent quantum system remains in its instantaneous eigenstate provided that its evolution is sufficiently slow compared with the relevant energy scales~\cite{Born1928Adiabatic,Bergmann1998CPT,MenchonEnrich2016SAP,Albash2018AQC,Polkovnikov2005UAD,Tong2007Adiabatic,Comparat2009Adiabatic}. 
This principle has become a cornerstone in quantum control, enabling reliable state preparation and the implementation of quantum simulations. 
In recent years, analogies between quantum and classical systems have extended the scope of the adiabatic theorem, finding applications in electron transport~\cite{vanWees1991Adiabatic,EntinWohlman2002Adiabatic,Rosso2003Electron}, wave scattering in complex media~\cite{Chase1956Adiabatic}, topological phenomena~\cite{Lahini2008Nonlinearity,Hamma2008Topological,yves2017topological,yves2017crystalline}, and the design of functional metamaterials~\cite{zhu2016implementation,Shen2019Acoustic,nassar2020nonreciprocity}. 
In both quantum and classical settings, the presence of dissipation or environmental coupling motivates the need for adiabatic processes that are as fast as possible. 
When the system evolves too rapidly, however, adiabaticity breaks down and nonadiabatic transitions (NATs) between different eigenstates become significant. 
Considerable attention has therefore been devoted to restoring adiabatic behavior within a short evolution time through approaches collectively known as shortcuts to adiabaticity (STA)~\cite{GueryOdelin2019STA,Taras2021STA,bergmann2019roadmap}.

Several approaches to STA have been developed, including counterdiabatic driving~\cite{Demirplak2003Adiabatic,Berry2009Transitionless,Chen2010Shortcut,Tseng2012FastMode,Paul2015Shortcut}, 
fast quasi-adiabatic driving (FAQUAD)~\cite{MartinezGaraot2017Quasiadiabatic,Chung2019STA,Hung2019Quasiadiabatic}, 
invariance-based engineering~\cite{Lewis1969Invariant,Chen2010Frictionless,Chen2011Transport,Benseny2017Geometric}, 
and non-Hermitian shortcuts~\cite{Torosov2013NonHermitian,Torosov2014NHSRAP,Riva2021Adiabatic}. 
These methods have been successfully applied not only to quantum systems but also to classical analogs, 
enabling the realization of compact functional devices such as mode converters~\cite{Lin2012ModeConversion,Tseng2012Engineering}, 
Y-junctions~\cite{Torrontegui2013Splitting,MartinezGaraot2014ModeSorting}, directional couplers~\cite{Tseng2013Counterdiabatic,Tseng2014Directional,Tseng2014Robust,Chen2016Couplers,tang2022functional}, 
and polarization rotators~\cite{Chen2014Rotators}. 
Among the various STA strategies, the FAQUAD approach is particularly practical because it accelerates the adiabatic process while remaining within experimentally accessible parameter constraints. 
By reducing the rate of change of the time-dependent parameters near level crossings and increasing it elsewhere—a form of velocity optimization—FAQUAD keeps the deviation from adiabaticity, quantified by the so-called adiabaticity parameter, at a small value. 
As a result, NATs are strongly suppressed even in relatively rapid evolutions.

In this study, we extend the FAQUAD framework by introducing path optimization and demonstrate its effectiveness on an elastic-wave platform that allows direct visualization of the entire process. 
The system consists of two parallel elastic waveguides connected by bridges, where the thicknesses of the waveguides and bridges are tuned to control the detuning of the propagation constants~($\delta$) and the coupling strength~($\kappa$). 
Although the most straightforward approach to achieving energy transfer from the initial system to the target one is to vary these two parameters adiabatically along a straight path in the $(\kappa,\delta)$ parameter space, such a constraint is not necessary. 
Alternative paths through the parameter space can yield the same final state within a shorter propagation length, a concept we refer to as path optimization in addition to the velocity optimization in FAQUAD. 
Our analytical, numerical, and experimental results confirm the validity of this approach, which provides a practical route for designing metamaterials that manipulate elastic wave propagation in compact devices.

\section{Hamiltonian extraction and band structure}

\begin{figure}[t]
  \centering
  \includegraphics[width=1.0\linewidth]{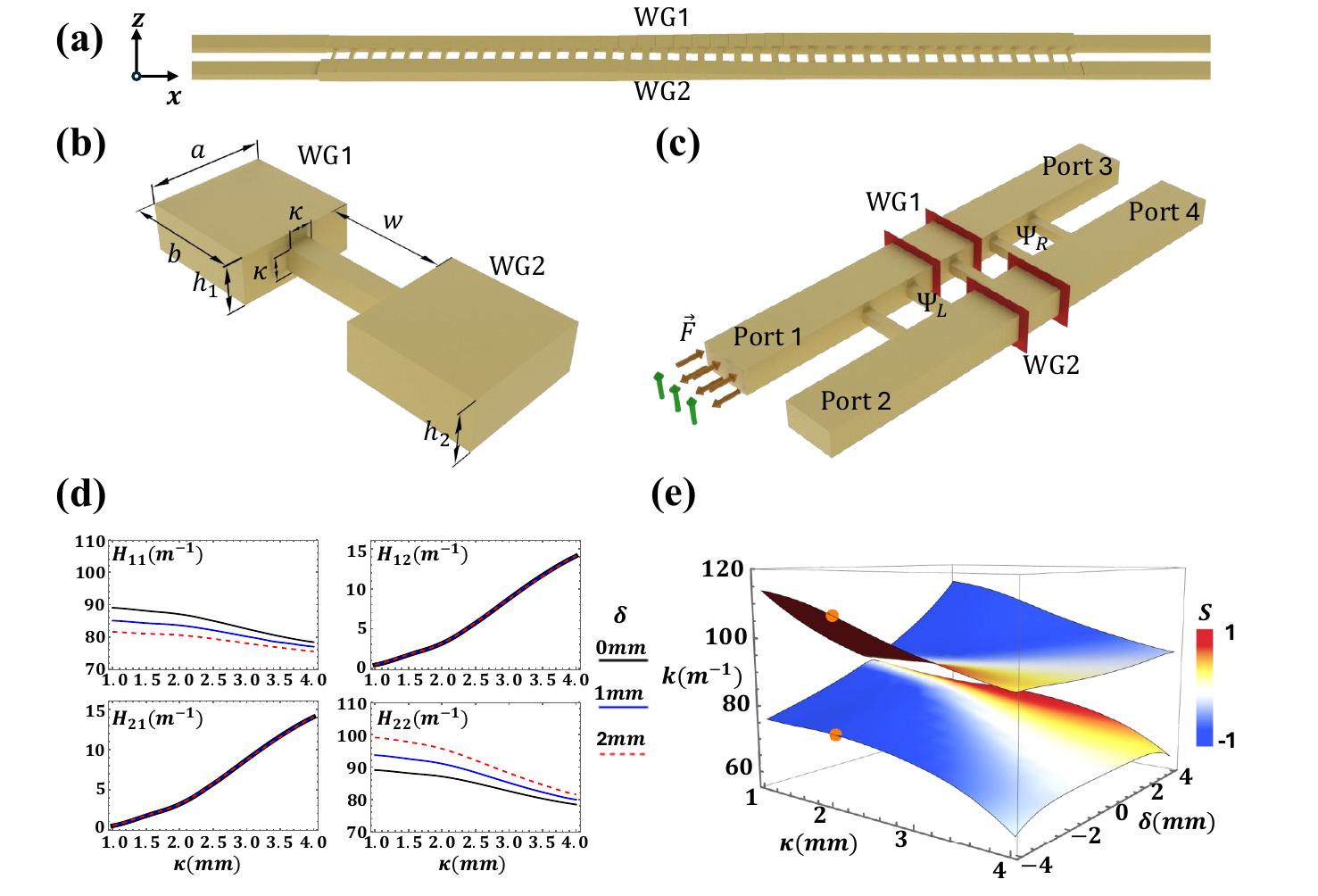}
  \caption{\textbf{Two-waveguide elastic system and band structure.} (a) Schematic of the two-waveguide system connected by elastic bridges. (b) Unit cell of the elastic system, where $h_1$ and $h_2$ are used to adjust the difference in the propagation constants between the two waveguides. (c) Procedure for extracting the transfer matrix $T$, with fields evaluated on both sides of the central unit (indicated by the red planes). (d) Components of the effective $2\times2$ Hamiltonian $H_{ij}$ as functions of the coupling parameter $\kappa$ for several detuning values $\delta$, demonstrating the dependence of mode coupling on geometric and material variations. (e) Band structures in the two-dimensional parameter space $(\kappa,\delta)$, with orange dots denoting experimentally verified bands. Modal parity parameter $S$ is plotted as color map on the surfaces. In panels (b) and (d), $\kappa$ and $\delta$ refer to geometric tuning parameters (bridge width and thickness difference), whereas in (e) they represent the effective coupling and detuning extracted from the transfer matrix. The elastic parameters used in the COMSOL simulations are $f=3\,\mathrm{kHz}$, $\omega=2\pi f$, $a=b=w=10\,\mathrm{mm}$, cross-sectional area $A=bh$, second moment-of-area $I=bh^3/12$, Young's modulus $E=3.4\,\mathrm{GPa}$, density $\rho=1190\,\mathrm{kg/m^3}$, shear modulus $\mu=0.35$, and longitudinal sound-speed $c=\sqrt{E/(12\rho)}$. Quantities with subscript zero correspond to the background double-waveguide (without bridges), where $h_0=4\,\mathrm{mm}$ and all other parameters remain unchanged.}
  \label{fig:fig1}
\end{figure}

These STA approaches were originally developed for quantum systems governed by the Schrödinger equation. 
To adapt them to elastic waveguides, we first examine the mathematical similarities between the elastic wave equation and the Schrödinger equation. 
In the elastic waveguide system depicted in Fig.~\ref{fig:fig1}(a), although the structure is three-dimensional, its analytical description can be effectively reduced to one dimension. 
This simplification arises because we focus on the flexural wave, which propagates along the $x$-direction while oscillating in the $z$-direction. 
The corresponding wave equation therefore takes a form analogous to the Schrödinger equation,
\begin{equation}
    -i\,\partial_x \psi = H \psi,
    \label{eq:propagation}
\end{equation}
where the propagation distance $x$ plays a role analogous to time in quantum dynamics. 
Here $\psi$ is a two-component state vector whose elements represent the complex amplitudes of the flexural waves on the two coupled waveguides, and $H$ is the corresponding $2\times2$ Hamiltonian governing their evolution.

Within this framework, our goal is to construct an equivalent Hamiltonian description that captures the essential coupling between the two elastic waveguides. 
Specifically, the Hamiltonian $H$ should satisfy the conditions required for implementing FAQUAD with two-dimensional parameter-space optimization. 
To define the adiabaticity parameter, a minimum of two interacting modes is required; hence, we model the two elastic waveguides in Fig.~\ref{fig:fig1}(a) as two discrete energy levels in the quantum analogy. 
In this correspondence, the two energy levels in a time-dependent quantum system are represented here by two elastic waveguide modes with distinct propagation constants, while the interaction between them arises from the coupling bridges.
The diagonal elements of $H$ represent the propagation constants of the flexural modes supported by each waveguide, while the off-diagonal terms describe the inter-waveguide coupling mediated by the connecting bridges. 
As illustrated in Fig.~\ref{fig:fig1}(b), the thickness difference between waveguide~1 (WG1) and waveguide~2 (WG2), defined as $\delta = h_1 - h_2$ while keeping $h_1 + h_2 = 10mm$ constant, controls the detuning between their propagation constants, whereas the cross-sectional area of the bridges, proportional to $\kappa^2$, determines the coupling strength. 
These two parameters, $\kappa$ and $\delta$, thus define the accessible parameter space for engineering the system’s Hamiltonian. 
By varying the geometry from unit to unit along the propagation direction, as shown in Fig.~\ref{fig:fig1}(a), we realize a spatially varying Hamiltonian that emulates temporal driving in quantum systems.

\medskip
\noindent\textbf{Numerical extraction of the effective Hamiltonian.}
To extract the Hamiltonian for structures defined by specific geometries $(h_1,h_2,\kappa)$, 
we perform full-wave simulations (COMSOL Multiphysics) using a configuration of five identical unit cells, as illustrated in Fig.~\ref{fig:fig1}(c). 
The total length of five cells is chosen to be sufficiently large to capture the near-field interaction between adjacent units, 
so that the field in the central cell reflects an effective description of the bulk medium. 
Flexural waves are excited by either bending moments (brown arrows) or shear forces (green arrows) applied at one of the four ports, resulting in eight possible excitation configurations in total. 
For each configuration, the flexural field is integrated over the cross-sections on the left and right sides of the central unit cell (red planes in Fig.~\ref{fig:fig1}(c)) to obtain the averaged field quantities.

The extracted flexural field on either side can be organized into an $8\times8$ matrix. 
For example, the field matrix on the left side is expressed as
\begin{equation}
\Psi_L =
\begin{pmatrix}
w^{1,1} & w^{2,1} & \cdots & w^{8,1} \\
\partial_z u^{1,1} & \partial_z u^{2,1} & \cdots & \partial_z u^{8,1} \\
M_{xx}^{1,1} & M_{xx}^{2,1} & \cdots & M_{xx}^{8,1} \\
Q_{xz}^{1,1} & Q_{xz}^{2,1} & \cdots & Q_{xz}^{8,1} \\
w^{1,2} & w^{2,2} & \cdots & w^{8,2} \\
\partial_z u^{1,2} & \partial_z u^{2,2} & \cdots & \partial_z u^{8,2} \\
M_{xx}^{1,2} & M_{xx}^{2,2} & \cdots & M_{xx}^{8,2} \\
Q_{xz}^{1,2} & Q_{xz}^{2,2} & \cdots & Q_{xz}^{8,2}
\end{pmatrix},
\label{eq:PsiL}
\end{equation}
where the first superscript denotes the excitation index and the second the waveguide index.
For example, the first column corresponds to excitation by a shear force at port~1.
The upper four block rows correspond to fields extracted from WG1 and the lower four from WG2, 
each containing the transverse displacement $w$, rotation $\partial_z u$, bending moment 
$M_{xx} = \frac{1}{b}\!\int dA\, z\,\sigma_{xx}$, and shear force 
$Q_{xz} = \frac{1}{b}\!\int dA\,\sigma_{xz}$.

To facilitate model analysis, we actually normalize the field components into dimensionless quantities using the scaling
\[
\frac{\omega}{c}w,\quad \frac{\omega h_0}{c}\,\partial_z u,\quad
\frac{A_0}{E_0 I_0} M_{xx},\quad
\frac{A_0 h_0}{E_0 I_0} Q_{xz},
\]
to assemble the state vector and the normalization parameters are defined in the caption of Fig.~\ref{fig:fig1}. 
After normalization, the transfer matrix for a given geometry of unit cell is obtained directly as
\begin{equation}
    T = \Psi_R\, \Psi_L^{-1},
    \label{eq:transfer}
\end{equation}
and the corresponding Hamiltonian ($8\times 8$ here) is obtained by
\begin{equation}
    H = \frac{\ln(T)}{i a},
    \label{eq:HfromT}
\end{equation}
where $a$ is the unit-cell length.

We then transform the Hamiltonian from the field representation to the modal basis of a background system. 
Here, the background refers to the pair of waveguides of height $h_0 = 4~\mathrm{mm}$ without any connecting bridges. 
The coefficients for the state in terms of the model basis are denoted as
$\psi_f^1, \psi_b^1, \psi_{fe}^1, \psi_{be}^1, \psi_f^2, \psi_b^2, \psi_{fe}^2,$ and $\psi_{be}^2$. 
The superscripts~1 and~2 denote WG1 and WG2, while the subscripts $f$, $b$, $fe$, and $be$ represent forward-propagating, backward-propagating, forward-evanescent, and backward-evanescent modes, respectively. 
In this basis, the diagonal elements of the Hamiltonian correspond to the propagation constants of the individual modes, and the off-diagonal elements describe their mutual coupling.
In our system of unit cells, the evanescent modes decay rapidly within both waveguides, and backward scattering remains minimal because the geometries $(h_1,h_2,\kappa)$ vary gradually from site to site. 
We therefore neglect the evanescent components and assume negligible conversion between forward and backward propagating modes. 
Finally, we find the $2\times 2$ effective Hamiltonian using the two forward eigenmodes of the system of unit cells and only retaining the coefficients of the forward-propagating modes of the background basis (components $\psi_f^1$ and $\psi_f^2$), We put these two eigenvectors into columns of a matrix $U$ ($2\times 2$). We also write a diagonal matrix $K$ in storing the propagation constants of the two eigenmodes as diagonal values. Then, the effective $H$ matrix can be approximated by
\(
H = U K U^{-1} \approx U K U^{\dagger}.
\)
It serves as the model for analyzing the fast quasi-adiabatic dynamics with path optimization. Hereafter, we denote truncated state vector using 
\(
|\psi\rangle = 
\begin{pmatrix}
\psi_f^1,
\psi_f^2
\end{pmatrix}^T
\) and $H$ refers to the $2\times 2$ effective matrix, overloading the same symbol for the original $8\times 8$ Hamiltonian in Eq.~\eqref{eq:propagation} .

In Fig.~\ref{fig:fig1}(d), we show the numerically extracted matrix elements of the Hamiltonian $H$ in four panels. 
Each panel presents the dependence of the $2\times2$ Hamiltonian elements on the coupling parameter $\kappa$ for several values of $\delta$. 
From the results, we observe that the diagonal terms vary with both $\kappa$ and $\delta$, 
while the off-diagonal terms increase monotonically with $\kappa$ but remain nearly unchanged as $\delta$ varies. 
Hence, the off-diagonal components of $H$ (found to be real as a good approximation), representing the coupling between the two waveguides, 
can be tuned by adjusting the cross-sectional area of the connecting bridges proportional to $\kappa^2$. 
On this basis, the diagonal elements of $H$, which represent the propagation constants of the two modes (and whose difference corresponds to the detuning), 
can be further controlled by varying the thickness difference between the two waveguides. These results confirm our intuition to use $\kappa$ and $\delta$ to control the Hamiltonian.

Figure~\ref{fig:fig1}(e) presents the band structure, eigenvalues of $H$, as propagation constant $k$ in the parameter space $(\kappa,\delta)$ at a chosen operation frequency of $f = 3~\mathrm{kHz}$. 
The color of the bands represents the energy distribution between the two waveguides, quantified by the modal parity parameter
\begin{equation}
S = \frac{|\psi_f^1|^2 - |\psi_f^2|^2}{|\psi_f^1|^2 + |\psi_f^2|^2},
\label{eq:Sdefinition}
\end{equation}
where red (blue) regions correspond to the energy being predominantly concentrated in WG1 (WG2). 
We denote $k_1$ and $k_2$ as the propagation constants associated with the lower-band eigenmode $|\psi_1\rangle$ and the higher-band eigenmode $|\psi_2\rangle$, respectively. Here, each of these eigenmodes have two real components due to the real Hamiltonian.

\section{Enhanced FAQUAD with path optimization}

\begin{figure}[t]
  \centering
  \includegraphics[width=0.9\linewidth]{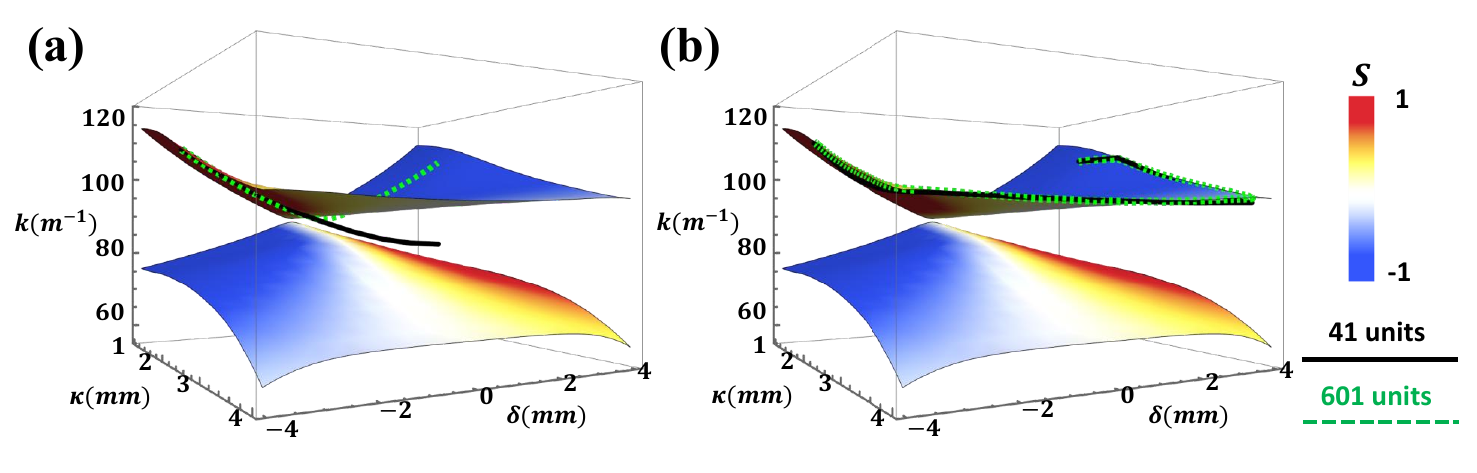}
  \caption{\textbf{Optimized paths in the parameter space.} (a,b) Paths in the $(\kappa,\delta)$ parameter space for the two designed paths. The black curves denote paths obtained from the analytic model for paths~1 and~2, respectively. The green curves correspond to the same paths executed over a much longer waveguide length (601 instead of 41 unit cells), representing the quasi-adiabatic limit. The colored surfaces show the band structure in $(\kappa,\delta)$, where the propagation constant $k$ acts as the eigenvalue analogue of energy in a time-dependent system. The color indicates the modal parity parameter $S$, quantifying the composition of the two hybridized modes along the path.
}
  \label{fig:fig2}
\end{figure}

\begin{figure}[t]
  \centering
  \includegraphics[width=0.9\linewidth]{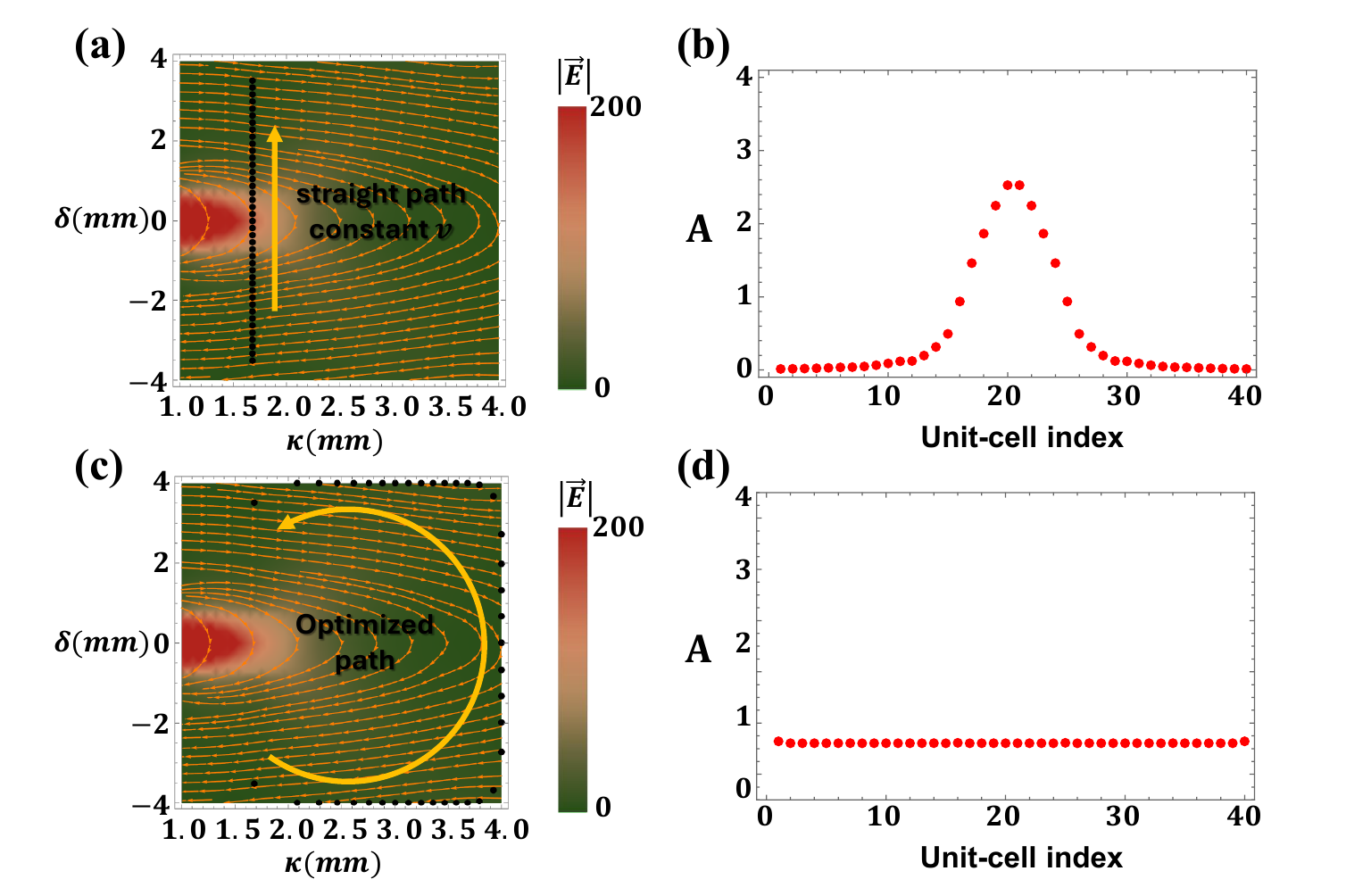}
  \caption{\textbf{Comparison between unoptimized and optimized paths.} (a,c) Vector field of $\vec{E}$ in the $(\kappa,\delta)$ parameter space, where color denotes the field intensity $|\vec{E}|$ and orange arrows indicate its local direction. The black dots represent the discrete path along which the system evolves: (a) a straight, constant-speed path (path~1) and (c) an optimized path with spatially varying speed (path~2).(b,d) Corresponding adiabaticity parameters in discrete form for the two paths, showing the local degree of adiabatic following between adjacent unit cells. The discrete adiabaticity parameter is obtained from Eq.~\ref{eq:adiabaticity} by replacing $\dot{\ket{\psi_1}}$ with $(\ket{\psi_1(i+1)}-\ket{\psi_1(i)})/a$, where $a$ is the unit-cell length.}
  \label{fig:fig3}
\end{figure}

\medskip
\noindent\textbf{Adiabatic vs. nonadiabatic evolution.}
After formulating the flexural-wave propagation in the double-waveguide system in a form analogous to the Schrödinger equation with a $2\times2$ Hamiltonian, 
we first examine how the system evolves under slow or rapid spatial modulation to distinguish adiabatic~\cite{Born1928Adiabatic,Bergmann1998CPT} from nonadiabatic behavior. 
To demonstrate this, we consider a controlled energy transfer from WG1 to WG2 by defining a path in the $(\kappa,\delta)$ parameter space that connects the initial and final configurations. 
Each unit cell of the waveguide is assigned a geometry corresponding to a point on this path, 
so that the local thicknesses and bridge widths vary smoothly along the propagation direction. 
The simplest realization is a straight path in the two-parameter space, along which $\delta$ changes linearly while $\kappa$ remains constant, 
as illustrated by the green dashed line in Fig.~\ref{fig:fig3}(a), which serves as our target path in parameter space. 
Here, a ``straight'' path means that the projection of the green dashed line onto the $(\kappa,\delta)$ plane is linear. 
At this stage, we also assume a constant ``velocity,'' i.e., $\delta$ varies linearly along the sequence of unit cells in the double-waveguide system.

The significance of this path in parameter space can be understood from the band structure in Fig.~\ref{fig:fig1}(e). 
Along the chosen path, the upper-band eigenstate, denoted $|\psi_2\rangle$, undergoes a continuous change in modal localization, 
as indicated by the color-coded modal parity parameter $S$ on the two bands. 
At the starting point $(\kappa_i,\delta_i)=(1.68,-3.51)$, 
the mode energy is concentrated in WG1 (red region on the band), 
while at the endpoint $(\kappa_f,\delta_f)=(1.68,3.51)$, 
it is concentrated in WG2 (blue region on the band). 
Therefore, if the system evolves adiabatically along this path, 
the energy initially confined to WG1 will gradually transfer to WG2 while remaining on the same upper-band eigenstate.

The effectiveness of this process depends on the spatial length available for adiabatic evolution. 
Figure~\ref{fig:fig2}(a) compares the results for two double-waveguide systems of different total lengths following the same straight path. 
For a 41-unit structure (41~cm), the state cannot remain on $|\psi_2\rangle$, 
indicating NATs between the bands. 
This can be visualized by tracing the measured propagation constant $k$ using
\[
\langle H \rangle
= \frac{\langle \psi | H | \psi \rangle}{\langle \psi | \psi \rangle},
\]
where the black curve in the 3D band-structure plot corresponds to the measured propagation constant lying between the two bands. 
In contrast, a longer 601-unit structure (601~cm) provides sufficient distance for gradual evolution, 
allowing the state to follow $|\psi_2\rangle$ adiabatically along the entire path 
(the green curve, where the measured $k$ remains on the upper band) and achieve complete energy transfer.

\medskip
\noindent\textbf{FAQUAD and path optimization.}
To quantify how slowly the system parameters must vary to satisfy the adiabatic condition, 
we introduce the adiabaticity parameter $A$~\cite{MartinezGaraot2017Quasiadiabatic,Chung2019STA,Hung2019Quasiadiabatic}, which measures the coupling between the eigenstates of the two bands, 
$|\psi_1\rangle$ and $|\psi_2\rangle$, with propagation constants $k_1$ and $k_2$ along the $x$-direction of the double waveguide:
\begin{equation}
A = \left| \frac{\langle \psi_2 | \partial_x \psi_1 \rangle}{k_2 - k_1} \right|.
\label{eq:adiabaticity}
\end{equation}
A small value $A \ll 1$ ensures adiabatic evolution, whereas larger values indicate NATs. 
The goal of shortcut-to-adiabaticity schemes, such as FAQUAD, is to minimize $A$ while achieving the same target state within a shorter propagation distance, thereby reducing both device footprint and material loss.

In our implementation, we extend the FAQUAD framework—typically formulated for a single control parameter—into a two-parameter space by introducing both path and velocity optimizations as a two-step procedure. 
The first step, path optimization, explores alternative paths in the $(\kappa,\delta)$ parameter space, beyond straight lines, to minimize the total accumulated adiabaticity. 
The integral of $A$ along the path can be expressed either in the physical coordinate $x$ or equivalently as a contour integral in parameter space:
\begin{equation}
\int_{x_i}^{x_f} A\,dx 
= \int_{x_i}^{x_f} |\vec{E}\!\cdot\!\vec{v}|\,dx
= \int_C |\vec{E}\!\cdot d\vec{p}|,
\label{eq:Aint}
\end{equation}
where vector $\vec{p}(x) = (\kappa(x),\delta(x))$ defines the path $C$ in parameter space, 
$\vec{v} = d\vec{p}/dx = (d\kappa/dx,\, d\delta/dx)$ is the parameter “velocity,” 
and the pseudo-field $\vec{E}$ is defined as
\begin{equation}
\vec{E} =
\left(
\frac{\langle \psi_2 | \partial_\kappa \psi_1 \rangle}{k_2 - k_1},
\frac{\langle \psi_2 | \partial_\delta \psi_1 \rangle}{k_2 - k_1}
\right).
\label{eq:pseudofield}
\end{equation}
Figures~\ref{fig:fig3}(a) and \ref{fig:fig3}(c) visualize $\vec{E}$ in the $(\kappa,\delta)$ plane, 
where the color indicates $|\vec{E}|$ and the orange arrows show its direction. 
Equation~\eqref{eq:Aint} reformulates the integral of $A$ as a contour integral in parameter space, 
where $\vec{E}$ acts as a given vector field. 
Minimizing $\int A\,dx$ therefore reduces to finding the optimal path in parameter space without explicitly specifying the velocity $\vec{v}$. We note that the definition of $\vec{E}$ has a sign flip if we multiply $-1$ to either of the eigenbasis from the real Hamiltonian. It will affect the direction of the $\vec{E}$ but it does not affect the integral of $A$ in Eq.~\ref{eq:Aint}.

Intuitively, $A$ accumulates more slowly when the path follows directions nearly orthogonal to $\vec{E}$ (where $\vec{E}\!\cdot\!d\vec{p}\!\approx\!0$), 
or passes through regions with larger band gaps, since $A\!\sim\!1/(k_2-k_1)$. 
When the accessible region in parameter space is confined to $\kappa \in [1,4]~\mathrm{mm}$ and $\delta \in [-4,4]~\mathrm{mm}$, 
the optimized path satisfying these criteria tends to move along the boundary of this region, as shown in Fig.~\ref{fig:fig3}(c) as black dots. 
The optimized path is actually numerically obtained by discretizing the parameter space, evaluating $\int A\,dx$ between grid points, 
and applying the Dijkstra algorithm~\cite{deng2012fuzzy} to find the path with minimum accumulated adiabaticity.

After determining the optimized path, we proceed to optimize the velocity $\vec{v}(x)$ along it, 
as the choice of $\vec{v}$ affects the instantaneous value of $A(x)$ at each position. 
Figure~\ref{fig:fig3}(b) shows the discrete $A(x)$ profile for the unoptimized straight path of Fig.~\ref{fig:fig3}(a) under constant velocity, 
where a pronounced peak of $A$ appears at the midpoint of the path. 
The FAQUAD method mitigates this by adapting $|\vec{v}(x)|$: 
it slows down where the pseudo-field is strong (i.e., the band gap is small) and speeds up where it is weak (large band gap), 
thereby maintaining a nearly uniform adiabaticity~\cite{MartinezGaraot2017Quasiadiabatic,Chung2019STA,Hung2019Quasiadiabatic}. 
This corresponds to enforcing
\begin{equation}
A(x) =  
\frac{\int_{x_i}^{x_f} A\,dx}{x_f - x_i},
\label{eq:FAQUAD}
\end{equation}
which defines the optimized velocity profile along the chosen path to give the constant averaged adiabaticity value.

Figure~\ref{fig:fig3}(d) shows the adiabaticity parameter $A(x)$ along the optimized path for a 41-unit structure. 
First, its average value is significantly smaller than that of the straight path, owing to path optimization. 
Second, it remains nearly uniform along the propagation direction as a result of velocity optimization, 
thereby avoiding the pronounced peak that exceeds the average value in the unoptimized case. 
Consequently, this configuration satisfies the adiabatic condition $A(x) \ll 1$ throughout the entire process. 
The black and green curves in Fig.~\ref{fig:fig2}(b) show the model-predicted propagation for different total lengths (41 and 601 units) along the optimized path. 
Even for the 41-unit case, the state evolves almost adiabatically along the upper band.

\section{Experimental validation}

\begin{figure}[t]
  \centering
  \includegraphics[width=1.0\linewidth]{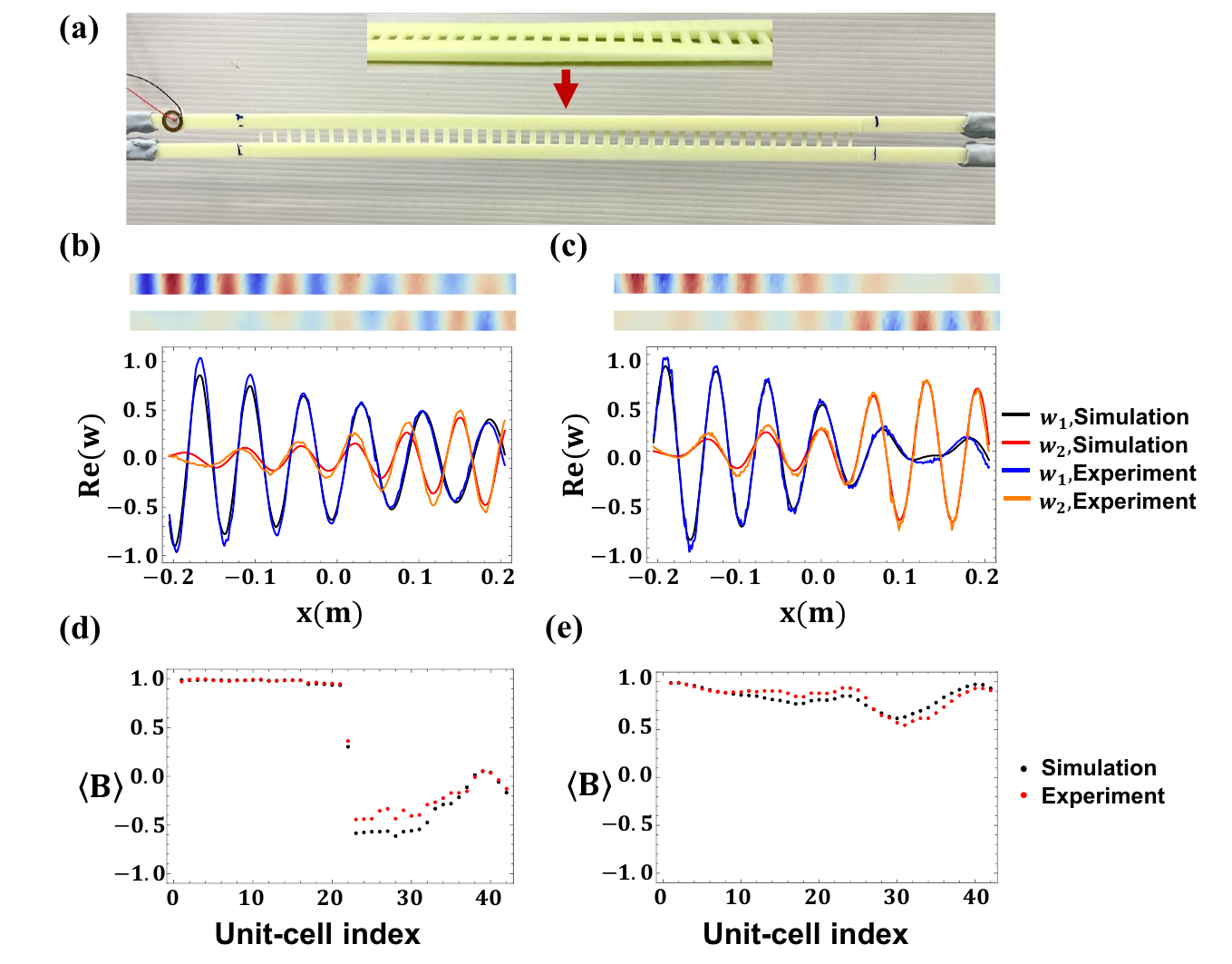}
  \caption{\textbf{Experimental validation of the optimized and unoptimized paths.} (a) Photograph of the experimental setup. The upper (lower) beam corresponds to WG~1 (2).  Blue tack is attached at both ends to serve as a perfect matching layer for suppressing reflections. A piezoelectric transducer is coupled to WG~1 to generate the prescribed input excitation. The structure contains 41 unit cells, with unit~1 located on the left and unit~41 on the right. The inset provides a side view of the fabricated sample. (b,c) Field distributions obtained from simulations and experiments for the paths shown in Fig.~2(a) and Fig.~2(b), respectively. The color maps represent the measured displacement fields, while the curves show the normalized real parts of the wave amplitudes $W_1$ and $W_2$ for the two coupled waveguides, comparing simulation (black and red) and experiment (blue and orange). (d,e) Corresponding band indices $\langle B \rangle$ extracted from simulations (black dots) and experiments (red dots) for the two paths, showing consistent evolution of modal character along the propagation direction.
}
  \label{fig:fig4}
\end{figure}

We next perform experiments to verify the effectiveness of the enhanced FAQUAD approach with path optimization. 
The experimental setup is shown in Fig.~\ref{fig:fig4}(a). 
The elastic double-waveguide sample is mounted on a support shelf, with blue tack applied to all four ends to act as a perfect matching layer that suppresses reflected waves. 
A piezoelectric transducer is attached to the left-hand side of the upper waveguide (WG1) to excite a flexural wave at 3~kHz, generating the predetermined input. 
A laser Doppler vibrometer  (Optomet SWIR Scanning Vibrometer) is then used to scan the amplitude and phase of the flexural vibration field along the structure on both $x$ and $y$ directions. 

Figures~\ref{fig:fig4}(b) and \ref{fig:fig4}(c) present the measured flexural fields for the straight and optimized paths (in parameter space), respectively. 
Two different samples, each consisting of 41 unit cells, are tested. 
In each case, the two-dimensional color map shows the real part of the displacement field, $\mathrm{Re}(w)$, on both waveguides, 
while the one-dimensional plots below display the field amplitude along the propagation direction ($x$), averaged across the transverse direction ($y$) of each waveguide to improve accuracy. 
Corresponding full-wave simulation results are also plotted for comparison. 
The close agreement between experiment and simulation confirms the validity of both the design and the measurement procedure. 

For the straight path [Fig.~\ref{fig:fig4}(b)], the oscillation amplitude initially concentrates in WG1 but remains distributed across both waveguides at the output end, 
indicating that the energy transfer from WG1 to WG2 is incomplete within this short propagation length without optimization. 
It is important to note that both the input and target output states correspond to the upper-band eigenmode $|\psi_2\rangle$, 
but they occupy different positions in the $(\kappa,\delta)$ parameter space. 
In contrast, for the optimized path [Fig.~\ref{fig:fig4}(c)], the oscillation amplitude starts in WG1 and successfully transfers to WG2 by the end of the structure, 
demonstrating that the enhanced FAQUAD approach with both path and velocity optimization achieves the desired adiabatic evolution and efficient energy transfer within a compact device length.

For a more quantitative analysis, we map the measured field patterns onto the eigenmode coefficients for both the straight and optimized paths. 
In the experimental data, the evanescent modes are neglected, and the measured state is decomposed into the forward-propagating modal components $\psi_f^1$ and $\psi_f^2$, 
defined with respect to the background double-waveguide. 
After obtaining the state 
\(
|\psi\rangle = 
\begin{pmatrix}
\psi_f^1,
\psi_f^2
\end{pmatrix}^T
\)
along the propagation direction, we evaluate its band index by rescaling the measured propagation constant to the local band-gap size:
\begin{equation}
\langle B \rangle = \frac{2\langle H \rangle -k_1-k_2}{k_2 - k_1},
\label{eq:bandindex}
\end{equation}
where $\langle B \rangle$ ranges from $-1$ to $1$: 
$\langle B \rangle = -1$ corresponds to the lower band, 
$\langle B \rangle = 1$ to the upper band, 
and intermediate values indicate a mixed state. 
The measaured propagation constant  $\langle H \rangle$ also enables us to track the path of the measured state in the two-dimensional parameter space.

Figures~\ref{fig:fig4}(d) and \ref{fig:fig4}(e) show the extracted band index at the interfaces between neighboring unit cells. 
Black and red dots represent the simulation and experimental results, respectively, which nearly overlap—confirming the reliability of the measurements. 
For the straight path, the band index varies from $+1$ to $-1$, indicating a breakdown of adiabaticity and strong interband transitions. 
In contrast, for the optimized path, $\langle B \rangle$ remains close to $+1$ throughout the propagation, 
demonstrating that the state stays on the upper band and evolves adiabatically even within a short structure. 
For comparison, the paths of the model-predicted $\langle H \rangle$ in Fig.~\ref{fig:fig2} exhibit the same behavior, 
further confirming that the enhanced FAQUAD scheme enables faithful adiabatic driving within a compact device length.

\section{Conclusion}

In this work, we have introduced an enhanced FAQUAD approach for realizing adiabatic driving through simultaneous path and velocity optimization. 
Unlike conventional FAQUAD schemes, which adjust only the parameter velocity along a predefined path, our method also optimizes the path itself in the two-dimensional parameter space. 
A key advantage of this framework is that the optimal path can be determined solely from the geometry of the parameter space and the band structure, 
without prior knowledge of the velocity profile, which can subsequently be optimized once the path has been established. 
By combining these two steps, the average adiabaticity parameter remains small even in compact devices, while its instantaneous value stays nearly uniform along the propagation direction, enhancing the possibility to restoring adiabatic evolution throughout the process. 

Using the proposed path optimization, we designed and experimentally demonstrated efficient energy transfer of flexural waves in coupled elastic waveguides—a task unattainable without such optimization. 
Both full-wave simulations and experimental measurements confirm the validity and robustness of this strategy. 
Compared with quantum and optical systems, the elastic-wave platform offers a unique advantage: the entire adiabatic process can be directly visualized, providing an ideal testbed for developing, verifying, and refining shortcut-to-adiabaticity schemes.

\medskip
\noindent\textbf{Acknowledgements}
This work was supported by the Research Grants Council (RGC) of Hong Kong through Grant No.~AoE/P-502/20 and 16307522.



\bibliography{bib}

\end{document}